\documentclass[aps,prl,superscriptaddress,twocolumn,floatfix,a4paper]{revtex4}

\usepackage{graphicx,graphics,epsfig}   
\usepackage{dcolumn}    
\usepackage{bm}         
\usepackage{amsmath}    
\usepackage{verbatim}   
\usepackage{color}      
\usepackage{subfigure}  
\usepackage{times,natbib}
\usepackage{amsmath,amsfonts,amssymb,graphics,graphicx,epsfig,color,times,natbib}

\newcommand{\be}{\begin{equation}}
\newcommand{\ee}{\end{equation}}
\newcommand{\ba}{\begin{eqnarray}}
\newcommand{\ea}{\end{eqnarray}}
\newcommand{\ban}{\begin{eqnarray*}}
\newcommand{\ean}{\end{eqnarray*}}

\begin{document}


\title{Recovering part of the quantum boundary from information causality}
\author{Jonathan Allcock}
\email{jon.allcock@bristol.ac.uk}
\affiliation{Department of Mathematics, University of Bristol, Bristol BS8 1TW,
United Kingdom}
\author{Nicolas Brunner}
\affiliation{ H.H. Wills Physics Laboratory, University of Bristol, Tyndall
Avenue, Bristol, BS8 1TL, United
Kingdom }
\author{Marcin Pawlowski}
\affiliation{Institute of Theoretical Physics and Astrophysics, University of
Gdansk, 80-952 Gdansk, Poland}
\author{Valerio Scarani}
\affiliation{Centre for Quantum Technologies and Department of Physics,
National University of Singapore, 3 Science Drive 2, 117543 Singapore,
Singapore}

\date{\today}


\begin{abstract}
Recently, the principle of information causality has appeared as a good
candidate for an information-theoretic principle that would single out quantum
correlations among more general non-signalling models. Here we present results
going in this direction; namely we show that part of the boundary of quantum
correlations actually emerges from information causality.
\end{abstract}

\maketitle

\section{I.\hspace{0.3 in}Introduction}

Non-locality is a central feature of quantum mechanics (QM) and a powerful resource
for processing information. However, as Tsirelson \cite{tsirelson2} first
proved, the amount of non-locality allowed by QM is limited. In a
seminal paper, Popescu and Rohrlich \cite{PR} showed that this limitation is not
a consequence of relativity. Indeed, there exist theories which are more
non-local than QM yet do not allow for superluminal signalling.
Identifying the physical principles underlying the limits to quantum
non-locality is now a central problem in foundational QM.

Recently, several works have studied the physical and information-theoretic
properties of general non-signalling models. Surprisingly, it appears that
these
models have numerous properties in common with QM, such as
no-cloning \cite{NS,john}, no-broadcasting \cite{barnum}, monogamy of
correlations \cite{NS} and information-disturbance trade-offs
\cite{praNScrypto}. General non-signalling models also allow for secure key
distribution \cite{BHK,NLcrypto} as well as quantum-like dynamical processes
\cite{emergence}. Therefore, none of these properties, usually thought of as
being typically quantum, are useful for separating quantum from post-quantum
correlations.

On the other hand, it is known that some particular post-quantum correlations
have extremely powerful communication properties. For instance, the availability
of PR boxes - the paradigmatic example of post-quantum correlations - makes
communication complexity trivial \cite{vanDam}. However, communication
complexity is not trivial in QM \cite{IP}, and it is strongly
believed not to be trivial in nature. Therefore correlations which collapse
communication complexity, such as PR box correlations, appear unlikely to exist.
More recently, a similar conclusion has been shown to hold for two classes of
noisy PR boxes \cite{brassard,Distillation}. However, there is a large class of
post-quantum correlations for which it is still unknown whether communication
complexity collapses or not.

In parallel, non-locality has also been studied from the point of view of
non-local computation \cite{noah}. Remarkably, here Tsirelson's bound (of
quantum non-locality) naturally appears, since all post-quantum correlations
violating this bound offer an advantage over classical and quantum correlations.
It is also known that part of the quantum boundary emerges from non-locality
swapping \cite{emergence,couplers} (an analogue of entanglement swapping),
although the origin of this connection is still not understood. Finally
Tsirelson's bound also appears in theories with relaxed uncertainty relations
\cite{verSteeg}.

More recently, Pawlowski et al. \cite{IC} have introduced a new physical
principle, the principle of information causality (IC), which is satisfied by
both classical and quantum correlations. The essence of IC is that the
communication of $m$ classical bits can cause a potential information gain of at
most $m$ bits. As is the case for non-local computation, Tsirelson's bound
naturally emerges, since all correlations exceeding Tsirelson's bound violate
the principle of IC. Therefore IC is a potential candidate for separating
quantum from post-quantum correlations. However, Tsirelson's bound identifies
only one point on the boundary of the set of quantum correlations. There are
also post-quantum correlations which lie below Tsirelson's bound. Thus, while
the emergence of Tsirelson's bound from IC is a remarkable feature, it is not
sufficient for singling out quantum correlations. More generally, one aims at
finding a principle underlying the full quantum boundary.

In the present paper, we show that part of the quantum boundary actually emerges
from IC. More precisely, we show that in two 2-dimensional slices of the
binary-input/binary-output non-signalling polytope, the IC criterion
analytically coincides with the quantum boundary.

The organisation of the paper is the following. In Section II we review the
geometrical approach to non-signalling correlations, while in Section III we
review IC. In Section IV, we study the link between IC and the quantum boundary.


\section{II.\hspace{0.3 in}Geometry of non-signalling boxes}

It will be convenient to describe bipartite non-signalling correlations in terms
of black boxes shared between two parties, Alice and Bob. Alice and Bob input
variables $x$ and $y$ at their ends of the box respectively, and receive outputs
$a$ and $b$.  The behaviour of a given correlation box is fully described by a
set of joint probabilities $P(ab|xy)$. We focus on the case of binary inputs
and outputs ($a,b,x,y\in\left\{0,1\right\}$), for which
\ba\nonumber  P(ab|xy)= \frac{1}{4} \left[ 1 + \left(-1\right)^{a}C_x +
\left(-1\right)^{b}C_y + \left(-1\right)^{a\oplus b}C_{xy} \right] \ea where
$\oplus$ is addition modulo 2, and
the correlators are given by
$C_{xy} = \sum_{a^{\prime} = b^{\prime}}P(a^{\prime}b^{\prime}|xy)- \sum_{a^{\prime}\neq b^{\prime}}P(a^{\prime}b^{\prime}|xy)$,
and the marginals by
$C_x = \sum_{b^{\prime}}\left[P(0b^{\prime} | x0) - P(1,b^{\prime} | x0)\right]$ and
$C_y= \sum_{a^{\prime}}\left[P(a^{\prime}0|0y) - P(a^{\prime}1 |0 y)\right]$.
 In this case, which corresponds to the famous Clauser-Horne-Shimony-Holt (CHSH)
\cite{chsh} scenario, the full set of non-signalling boxes forms an
8-dimensional polytope \cite{barrett} which has 24 vertices: 8 extremal
non-local boxes and 16 local deterministic boxes. The extremal non-local
correlations have the form:
\begin{equation*}	
P_{\text{NL}}^{\mu\nu\sigma}(ab|xy) =
	\begin{cases}
		\frac{1}{2} & \text{if} \quad \text{$a \oplus b = xy\oplus\mu
x\oplus\nu y\oplus\sigma $} \\
		0 & \text{otherwise}
	\end{cases}
\end{equation*}
where $\mu,\nu,\sigma\in\left\{0,1\right\}$, and the canonical PR box
corresponds to $\text{PR}= P_\text{NL}^{000}$. Similarly, the local
deterministic boxes are described by
\begin{equation*}	
P_{\text{L}}^{\mu\nu\sigma\tau}(ab|xy) =
	\begin{cases}
		1 & \text{if} \quad\text{$a = \mu x \oplus\nu$} \quad \text{$b =
\sigma y\oplus\tau$} \\
		0 & \text{otherwise}
	\end{cases}
\end{equation*}
The set of local boxes forms a subpolytope of the full non-signalling polytope,
and has facets which correspond to Bell inequalities - here the CHSH inequality
\begin{equation} \label{CHSH}
	C_{00} + C_{01} + C_{10}-C_{11} \leq 2,
\end{equation}
and its symmetries. Note that there are 8 symmetries of the CHSH inequality (any odd number of terms on the left hand side of \eqref{CHSH} can have a minus sign), and
that each CHSH inequality is violated by one of the extremal non-local boxes.

The set of quantum boxes, i.e. correlations obtainable by performing local
measurements on a quantum state (of any dimension), is sandwiched between the
local polytope and the full non-signalling polytope.  In particular, quantum correlations satisfy a variant of inequality \eqref{CHSH}, where the right hand side is replaced by $2\sqrt{2}$, a value known as Tsirelson's bound. The quantum set is a convex
body, although it is not a polytope. Thus, its boundary is described by a smooth
curve. For binary inputs and outputs, Tsirelson, Landau and Masanes (TLM)
\cite{TLM} have (independently) derived a necessary and
sufficient criterion for a set of correlators $C_{xy}$ to admit a quantum
description. In the form of Landau, $C_{xy}$ must satisfy:
\begin{equation} \label{TLM}
    |C_{00}C_{10}-C_{01}C_{11}|  \leq  \sum_{j = 0,1}\sqrt{(1-C_{0j}^2)(1-C_{1j}^2)}
\end{equation}
However, when considering the full probability distribution (including the
marginals), this criterion remains necessary but is no longer sufficient.
Recently, a refinement of \eqref{TLM} has been derived by Navascues, Pironio,
and Acin (NPA) \cite{miguel}. Their work improves \eqref{TLM} in that it
incorporates the marginals of the probability distribution. The NPA criterion
reads:
\ba	\label{NPA}
	| \text{asin} D_{00}+ \text{asin} D_{01}+ \text{asin} D_{10} -
\text{asin} D_{11} | \leq \pi,
\ea
where $D_{xy}= (C_{xy}-C_xC_y)/ \sqrt{(1-C_x^2)(1-C_y^2)}$. Note that for
vanishing marginals, \eqref{NPA} is equivalent to \eqref{TLM}. Note also that
\eqref{NPA} is in general not sufficient for a probability distribution to be
quantum-realizable; to determine whether a probability distribution is quantum
or not, one has to test a hierarchy of semi-definite programming conditions
\cite{SDP}.


\section{III.\hspace{0.3 in}Information Causality}

Let us now briefly review the principle of IC. The authors of
\cite{IC} considered the following communication task, which is similar to
random access coding \cite{ANTV02} and oblivious transfer \cite{OT, WW05}.
Alice and Bob, who are separated in space, have access to non-signalling
resources such as shared randomness, entanglement or (in principle) PR boxes.
Alice receives $N$ i.i.d. random bits $\vec{a}=\left(a_1,a_2,\ldots,a_N\right)$,
while Bob receives a random variable $b\in\left\{1,2,\ldots,N\right\}$.  Alice
then sends $m$ classical bits to Bob, who must output a single bit $\beta$ with
the aim of guessing the value of Alice's b-th bit $a_b$.  Their degree of
success at this task is measured by
\begin{equation*}
I \equiv\sum_{K=1}^{N}I\left(a_{K}:\beta|b=K\right),
\end{equation*}
where $I\left(a_{K}:\beta|b=K\right)$ is the Shannon mutual information between
$a_K$ and $\beta$. The principle of IC states that physically
allowed theories must have
$I \leq m$.
Indeed, it was proved in \cite{IC} that both classical and quantum correlations
satisfy this condition. Moreover, suppose that Alice and Bob share arbitrary
binary-input/binary-output non-signalling correlations corresponding to
conditional probabilities $P\left(ab|xy\right)$. A condition under which
IC is violated was derived in \cite{IC}  - based on a
construction by van Dam \cite{vanDam} and Wolf and Wullschleger \cite{WW05} -
for a specific realization of the Alice-Bob channel. It goes as follows. Define
$P_I$ and $P_{II}$:
\ba
P_{I} 	&= \frac{1}{2}\left[P(a\oplus b = 0 |00) + P(a\oplus b = 0|10)\right],
\notag \\
	&= \frac{1}{4} \left[2 + C_{00} + C_{10}\right] \notag \\
P_{II} 	&= \frac{1}{2}\left[P(a\oplus b = 0 |01) + P(a\oplus b = 1|11)\right].
\notag \\
	&= \frac{1}{4}\left[2 + C_{01} - C_{11}\right]. \label{PI}
\ea
Then, the IC condition 
($I \leq m$)
is violated for all boxes for which
\be	\label{ICviolation}
E_{I}^{2} + E_{II}^{2} > 1,
\ee
where $E_j = 2P_j - 1$. It follows from this that all non-signalling
correlations which violate Tsirelson's bound also violate IC.
To see this, it suffices to consider `isotropic' correlations of the form $\alpha \text{PR} + (1-\alpha) \openone$, where $\openone$ is the correlation box given by $P(ab|xy) = 1/4$ ($\forall a,b,x,y$).  For such boxes, $E_1 =E_2 = \alpha$ and \eqref{ICviolation} is satisfied when $\alpha > 1/\sqrt{2}$, which corresponds to violating Tsirelson's bound.
However, as
previously mentioned, there are correlations which lie below Tsirelson's bound
which are nonetheless unobtainable in QM. For such correlations,
it was not known whether the principle of IC singles out exactly those allowed
by quantum physics. We 
now offer a partial answer to this question.


\section{IV.\hspace{0.3 in}IC and the quantum boundary}

Here we
investigate the link between IC and the set of correlations
achievable in QM. We would like to determine whether the entire
quantum boundary can be recovered from the principle of IC. It will be
convenient to re-express the condition \eqref{ICviolation} for the violation of
IC in terms of the correlators $C_{xy}$:
\begin{equation} \label{ICviolation2}
	\left(C_{00} + C_{10}\right)^2 + \left(C_{01} - C_{11}\right)^2 > 4.
\end{equation}
Interestingly, this is equivalent to a violation of Uffink's quadratic inequality
\cite{Uff02}; note that Uffink's inequality is known to be strictly weaker than the TLM criterion \footnote{M. Seevinck, private communication; see also A. Cabello, Phys. Rev. A {\bf 72} 012113 (2005).}.

In the following, we compare \eqref{ICviolation2} with the TLM and NPA criteria for quantumness. We shall investigate several two-dimensional slices of the non-signalling polytope, which can be grouped into two families. More precisely, we consider noisy PR boxes of the form:
\begin{equation} \label{e:slice}
	\text{PR}_{\alpha,\beta} = \alpha\text{PR} + \beta\text{B} +
(1-\alpha-\beta) \openone,
\end{equation}
where B is an extremal non-local box in the first family, and an extremal local deterministic box in the second.
Remarkably, in the first family we find two different slices of the polytope where boxes that satisfy IC
coincide analytically with the set of quantum boxes. In other words, in these
slices IC exactly singles out quantum correlations in that all post-quantum
correlations violate IC. Note that because of the symmetry of the polytope, it
is sufficient here to focus on non-signalling boxes violating the CHSH
inequality \eqref{CHSH}, and not those which violate the 7 other symmetries of CHSH; basically, a non-signaling box can never violate more than one symmetry of CHSH.


\textbf{Family 1.} We first consider correlations of the form \eqref{e:slice}, where $\text{B} = P_{\text{NL}}^{\mu\nu\sigma}$, and $\mu\nu\sigma$ can take any values except for $000$ and $001$ (since these are co-linear with $\text{PR}$ and $\openone$). The corresponding correlators are given by $C_{00} = \alpha + (-1)^{\sigma}\beta$, $C_{01} = \alpha + (-1)^{\nu\oplus\sigma}\beta$, $C_{10} = \alpha + (-1)^{\mu\oplus\sigma}\beta$, and $C_{11} = -\alpha + (-1)^{\mu\oplus\nu\oplus\sigma\oplus 1}\beta$.

We see from \eqref{TLM} that if boxes of this form are
to be quantum-realizable then we require that $\alpha^2 + \beta^2 \leq
\frac{1}{2}$. Note that here the TLM criteria is necessary and sufficient for quantumness
since the probability distribution given by boxes in this family has a
specific form
\footnote{Tsirelson and Masanes \cite{TLM} proved that if
a given set of correlators satisfies the TLM criterion \eqref{TLM}, there exists
a probability distribution admitting a quantum description that has this set of
correlators. Here the question is whether this quantum probability
distribution could have the form of correlations in family 1. This is indeed the case since
any probability distribution can be brought to a form where the marginals are
unbiased and the joint probabilities satisfy
$P(a,b|x,y)=P(a\oplus1,b\oplus1|x,y)$. Importantly, this can be done by local
operations without modifying the value of the correlators; Alice and Bob share a
random bit $\alpha$ and subsequently modify their output according to
$a\rightarrow a\oplus \alpha$ and $b\rightarrow b\oplus \alpha$. Thus, the
TLM criterion is necessary and sufficient for quantumness in the slices of the
polytope in family 1.}.
On the other hand, we see from \eqref{ICviolation2} that if $\text{B} = \text{PR}_2 = P_{\text{NL}}^{010}$, then IC is violated when
\begin{equation}\label{IC_PR2}
	\alpha^2 + \beta^2 > \frac{1}{2}.
\end{equation}
Thus, in this particular slice of the
non-signalling polytope, a box violates IC if and only if it is post-quantum
(Fig. 1). Note that here we could have chosen $\text{B} =\bar{\text{PR}_2} = P_\text{NL}^{011}$ as well.

\begin{figure}[t]
  \includegraphics[width=0.98\columnwidth]{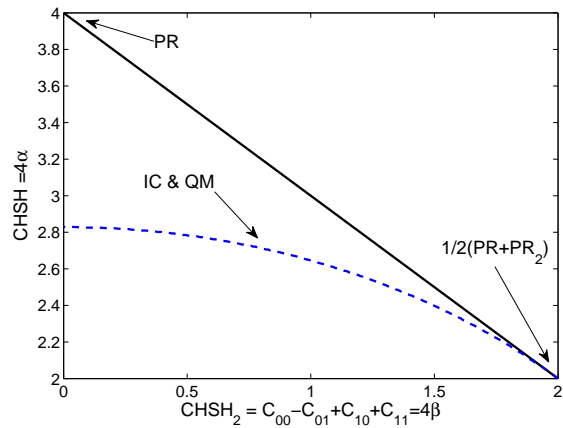}
  \caption{(Color online) A slice of the non-signalling polytope where correlations violate IC
if and only if they are post-quantum. Above the blue dashed curve, IC is
violated; below, correlations are quantum realizable.}
\end{figure}

The above proof is easily adapted to another slice. By exchanging the roles of Alice
and Bob, the same can also be seen to hold in the slice where $\text{B} =
\text{PR}_3 = P_\text{NL}^{100}$ (or equivalently $\text{B}=\bar{\text{PR}_3} = P_\text{NL}^{101}$).

Finally, note that in the case where $\text{B} = \text{PR}_4 =
P_\text{NL}^{111}$, the criterion for violating IC reduces to
$\alpha>\frac{1}{\sqrt{2}}$. Thus boxes below Tsirelson's bound are not known to violate IC in this slice (Fig. 2). We stress that this does not imply that there exist post-quantum boxes lying
below Tsirelson's bound which do not violate IC. The fact
that boxes which satisfy \eqref{ICviolation2} also violate IC follows from
considering a particular strategy for using the boxes, found in \cite{IC}. It
remains possible that a different strategy could be used to show that all
post-quantum correlations violate IC in this slice as well.

\begin{figure}[b]
  \includegraphics[width=0.98\columnwidth]{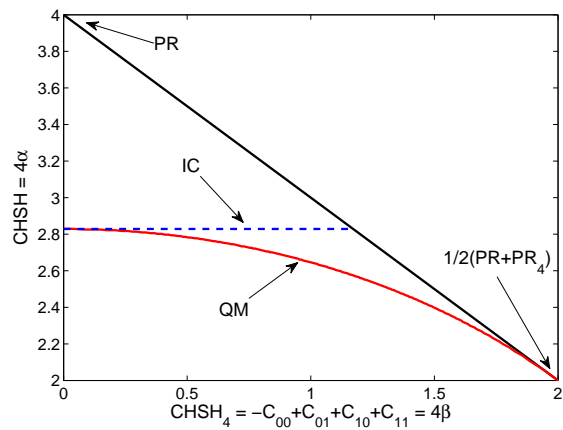}
  \caption{(Color online) A slice of the non-signalling polytope where post-quantum boxes which
lie below Tsirelson's bound ($\text{CHSH}=2\sqrt{2}$) are not known to violate
IC. The red solid line is the upper limit on quantum correlations, as given by
the NPA criteria.}
\end{figure}

\textbf{Family 2.} Next, we consider correlations of the form \eqref{e:slice},
where $\text{B} =  P_\text{L}^{\mu \nu \sigma \tau}$ with $\mu \sigma \oplus \nu \oplus \tau=0$; note that these are the local deterministic boxes sitting on the CHSH facet below the PR box. For simplicity, we will focus here on $B= P_\text{L}^{0000}$. In this case,
the correlators are given by $C_{00}=C_{01}=C_{10}=\alpha+\beta$,
$C_{11}=\beta-\alpha$, and the marginals by $C^a_0=C^a_1=C^b_0=C^b_1=\beta$. It
follows from \eqref{ICviolation2} that IC is violated whenever
\begin{equation} \label{IC_P0}
	(\alpha+\beta)^2 + \alpha^2 >1\,.
\end{equation}
However, this does not coincide with the NPA criterion \eqref{NPA}. Fig. 3 shows
clearly the discrepancy between the quantum boundary, or more precisely the
upper bound given by NPA, and the IC condition \eqref{IC_P0}. Let us re-iterate that the bound \eqref{IC_P0} follows from a particular
strategy in \cite{IC} for using boxes to violate IC. Thus it might still be the
case that a better strategy would single out quantum correlations in this
particular slice.

\begin{figure}[t]
  \includegraphics[width=0.98\columnwidth]{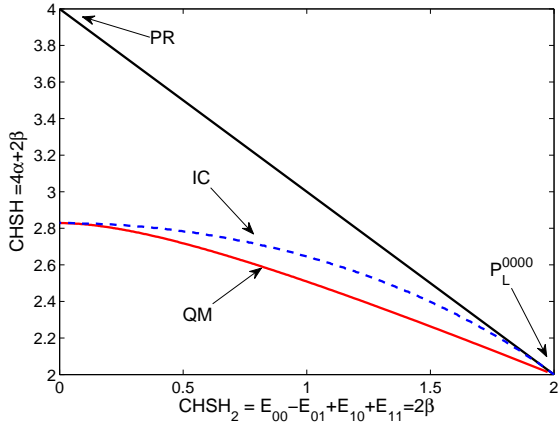}\\
  \caption{(Color online) A slice of the non-signalling polytope where IC does not single out
quantum correlations.}
\end{figure}


\section{V.\hspace{0.3 in}Conclusion}

We have shown that in the binary-input/binary-output non-signalling polytope,
part of the quantum boundary emerges from the principle of IC.  The central question is now whether this connection can be extended
to the full non-signalling polytope, which would establish IC
as the information theoretic principle singling out quantum correlations.

\emph{Acknowledgements.} We thank M.~Seevinck for pointing out the equivalence
of equation \eqref{ICviolation2} to Uffink's inequality. J.A. acknowledges support
from the Dorothy Hodgkins Foundation, N.B. from the Swiss National Science Foundation, M.P. from the EU
program QAP, and V.S. from the National Research
Foundation and the Ministry of Education, Singapore.

\bibliographystyle{prsty}
\bibliography{C:/BIB/thesis}

\end{document}